\documentclass{PoS}

\title{Microphysics in GRB central engine}

\ShortTitle{Microphysics in GRB central engine}

\author{\speaker{Agnieszka Janiuk}%
  \\
        Center for Theoretical Physics\\
        Polish Academy of Sciences \\
        Al. Lotnikow 32/46, 02-668 Warsaw \\
        E-mail: \email{agnes@cft.edu.pl}}

\author{Katarzyna Wojczuk and Konstantinos Sapountzis \\
        Center for Theoretical Physics, PAS\\
}
\abstract{We study the structure and evolution of the accreting plasma in gamma ray burst central engines. 
The models are based on the general relativistic MHD 
simulations. The nuclear equation of state adequate for dense and degenerate 
plasma, is incorporated to the numerical scheme instead of a simple polytropic 
EOS. Plasma is cooled by neutrinos and energy is extracted from the rotating 
BH by magnetic fields. We discuss our results in the frame of the observable
 GRBs, and speculate about the origin of variability in the GRB jet emission.
We also discuss the possible contribution from the heavy elements synthesized 
in winds from GRB engines to the kilonova emission.}

\FullConference{XII Multifrequency Behaviour of High Energy Cosmic Sources Workshop\\
                 12-17 June\\
                 Palermo, Italy}

\begin{document}

\section{Introduction}

The observed gamma ray bursts, which are generally grouped into two categries, namely
the long and the short events, present a large variety of time profiles and variability
of emission. Some of them consist of multiple peaks, while some present only a single
fast rise and exponential decay in the prompt gamma ray lightcurve. Also, the 
energetics involved in the explosions, albeit it is enormously large in comparison with any 
other known cosmic events, may differ by several orders of magnitude depending on the burst, 
which is inferred by their fluence detected with gamma-ray instruments.

There are several unknown factors that might be responsible for this variety, 
which are intrinsically related to the physics of jets, 
such as the efficiency of gamma ray production in the internal shocks, or the efficiency of
conversion between accretion and jet powers. Nevertheless, it is the accretion onto a 
black hole, which acts here and drives the explosion, 
with the gravitational potential energy being the most efficient mechanism of energy generation 
that we know.

To build a satisfactory model of the gamma ray burst central engine, 
like in all other cases where accretion processes are studied, we need to solve
the set of basic equations of hydrodynamics: the continuity, energy and momentum 
conservation equations. These can be treated with various simplifying assumptions, such as the
spherical or axial symmetry, or the prescription for the angular momentum transport with 
a so-called $\alpha$-disk model, which mimics the MHD turbulence.
We can also work out a stationary solution, before solving for a time-dependent evolution of 
the flow. All of these assumptions did work reasonably well in the context of accreting black 
holes in persistent sources, such as X-ray binaries or centers of active galaxies.
They were also used in the early works devoted to the GRB central engine modeling, and,
to some extent, are still used in the GRB context.

What actually makes a difference between the black hole accretion process 
that occurs in GRB engines and
all the other accreting black holes, is the microphysics of the plasma.
Due to the extremely transient nature of events, and hence
a huge accretion rate (which makes
about one Solar mass to be swallowed by the black hole in a timescale of one second),
the gas is very dense and hot, so that the onset the nuclear reactions occurs. 
Under the condition of $\beta$-equilibrium, the processes of electron and positron 
capture on nucleons, or electron-positron pair annihillation, lead to the efficient production
of neutrinos. These neutrinos carry the energy and provide a cooling mechanism to the disk
plasma, which is otherwise completely opaque to photons.
Also, heavier elements, starting from Helium nuclei, are synthesized within the disk, as 
well as in the winds launched from its surface.
These winds, possibly supported by the centrifugal force, but mainly magnetically- and neutrino- driven, 
may then
provide a site for nucleosyntesis of heavy nuclei via the r-process. 
In this way, they provide a chemical enrichment to the circum-burst medium, and they may 
act as a source of power to the emission of 'kilonovae', which are driven by the 
radioactive decay of unstable nuclei.

\section{Relativistic MHD model of gamma ray burst engine activity}

We based our simulations on the High Accuracy Relativistic Magnetohydrodynamics  code {\it HARM} \cite{Gammie03}
which provides solver for continuity and energy-momentum conservation equations in GR:
\[ \nabla_{\mu}(\rho u^{\mu}) = 0 ~~~~~ \nabla_{\mu}T^{\mu\nu} = 0 \]
where the energy tensor contains electromagnetic and gas parts:
\[ T^{\mu\nu} = T^{\mu\nu}_{\rm gas} + T^{\mu\nu}_{EM} \]
\[ T^{\mu\nu}_{gas} = \rho h u^{\mu} u^{\nu} + pg^{\mu\nu} =(\rho + u + p) u^{\mu} u^{\nu} + pg^{\mu\nu} \] 
\[ T^{\mu\nu}_{EM} = b^{2} u^{\mu} u^{\nu} + \frac{1}{2}b^2 g^{\mu\nu} - b^{\mu} b^{\nu} ; ~~ b^{\mu} = u_{\nu}{^{^*}\!\!F}^{\mu\nu} \]
Here $u^\mu$ is four-velocity of gas, $u$ is internal energy density, and 
$b^\mu = \frac{1}{2} \epsilon^{\mu\nu\rho\sigma}u_\nu F_{\rho\sigma}$
and $F$ is the electromagnetic stress tensor.
Note that in the force-free approximation, $E_{\nu}=u_{\mu}F^{\mu\nu}=0$.

\begin{figure}
\includegraphics[width=3.5cm]{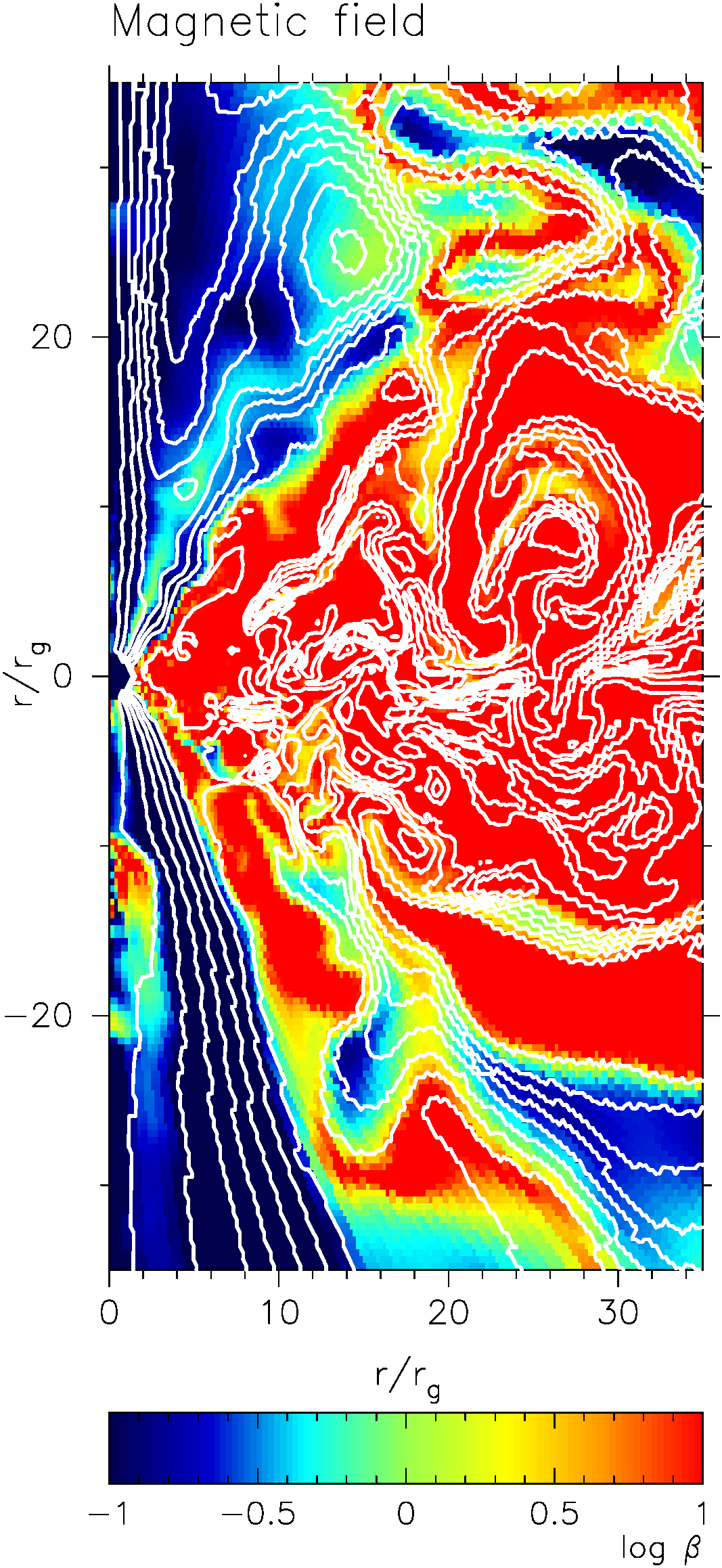}
\includegraphics[width=3.5cm]{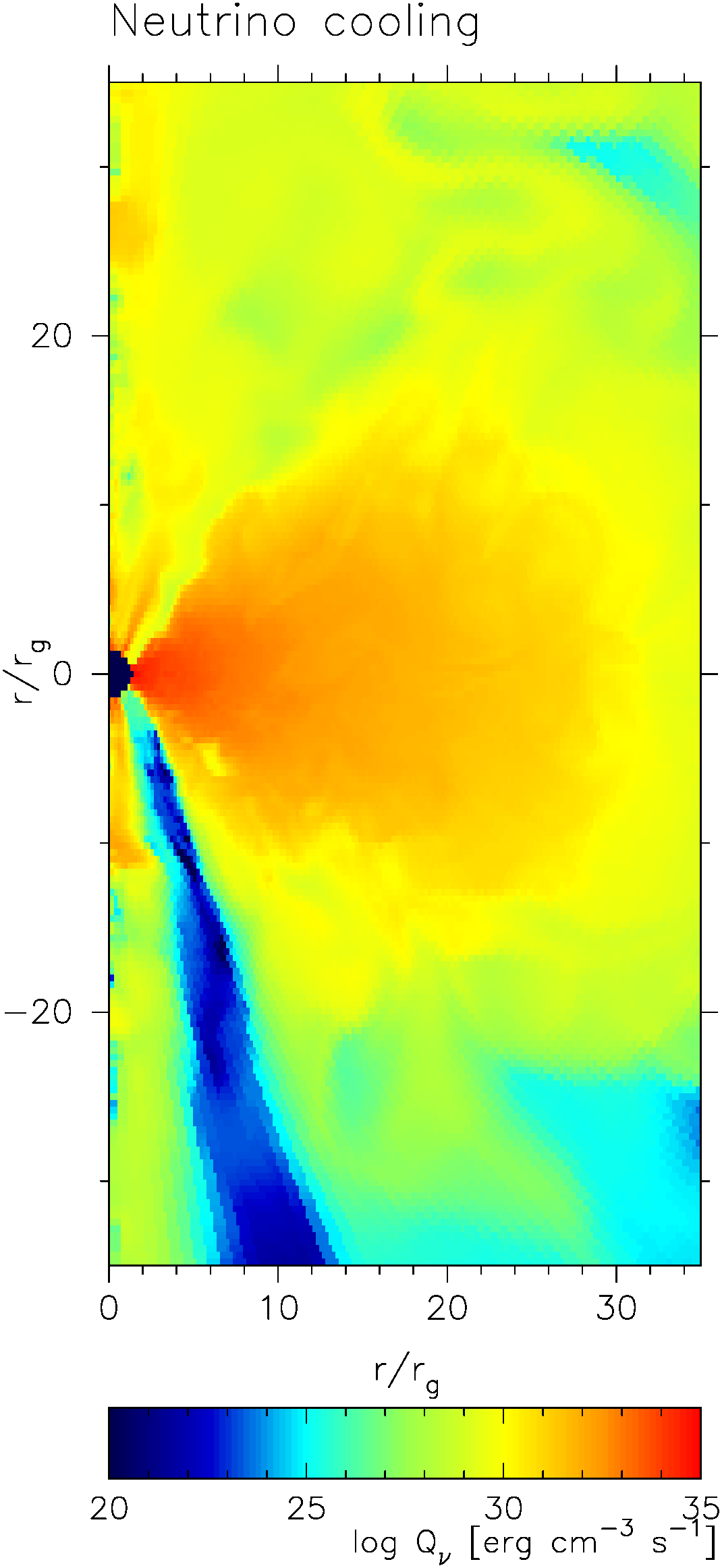}
\includegraphics[width=3.5cm]{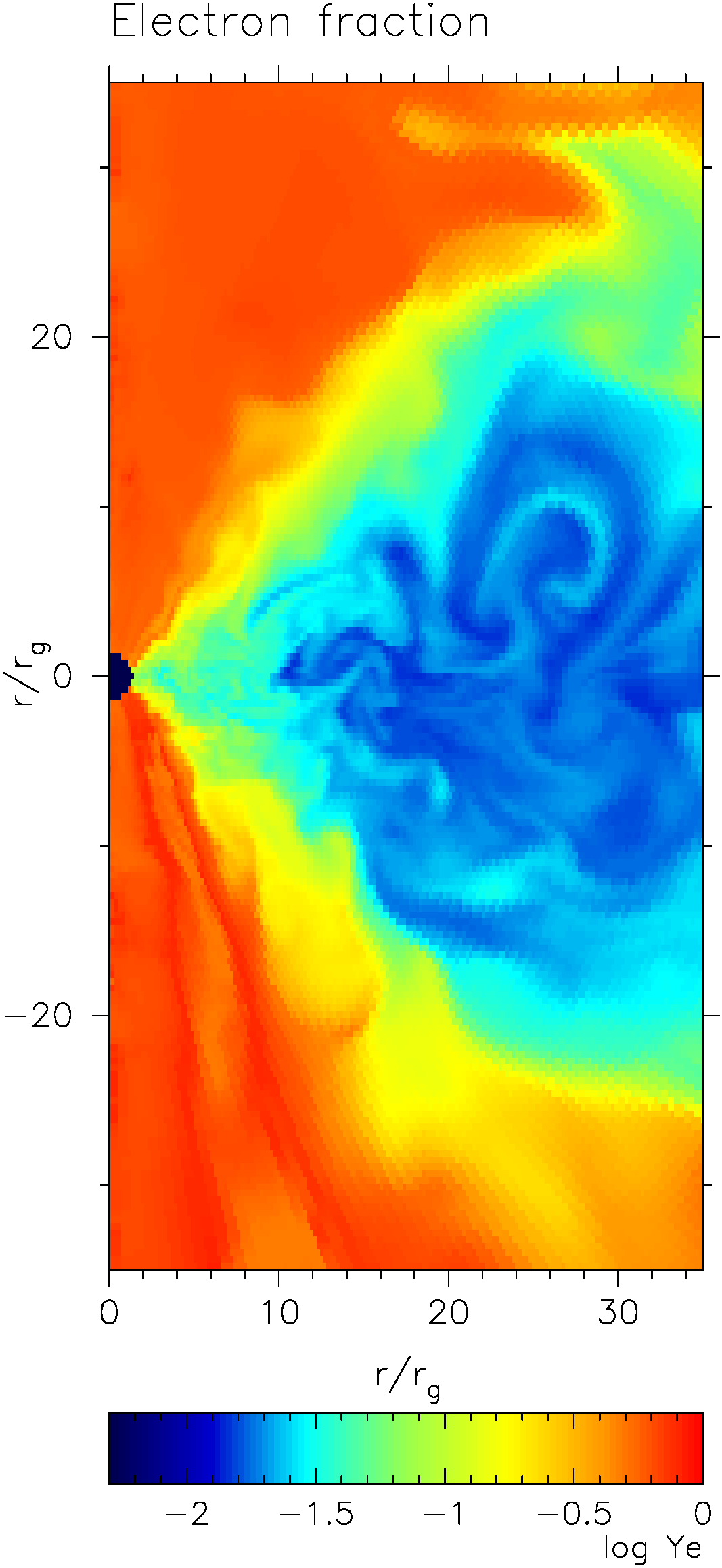}
\caption{Distribution of magnetic field lines, the gas to magnetic pressure ratio (left), 
neutrino emissivity (middle), and electron fraction (right), in the innermost 40 gravitational radii from the black hole. 
The evolved state of the flow is plotted at time t=2000 M, after the initial conditions have been relaxed.
This model global parameters adopt the black hole mass of 3 $M_{\odot}$, its spin of $a=0.98$, and disk mass is $0.1 M_{\odot}$.}
\label{fig:maps}
\end{figure}

In the simulations presented here 
in Fig. \ref{fig:maps}, the assumed 
initial condition invoked the relativistic equilibrium torus in the 
potential of a Kerr black hole, as described by \cite{fishbone}.
In addition, the initial seed 
poloidal magnetic field, with $A_{\phi}= (\rho/\rho_{max})$, was imposed, and normalized with $\beta=P_{gas}/P_{mag}=50$.

We observe that the field is advected with gas under the BH horizon
and close to the poles, the mass density is low
while the magnetic pressure is high.
The black hole rotation helps launching the GRB jets in two ways.

The mass-energy is dragged with the flow under the black hole horizon, and it
will provide a source of power to the relativistic jets, which 
are accelerated to the ultrarelativistic speeds.
At large distances from the engine, they become transparent to photons and
are the sites of gamma ray production.

We can evaluate the radial energy flux that is threading the black hole horizon, 
and define in this way the power of the Blandford-Znajek process:
\[ \dot{E} \equiv 2\pi\int_0^{\pi} \, d\theta \,
{\sqrt{-g}}{F_E}  \]
where $F_E \equiv -{{T^{r}_t}}$.  
This can be subdivided into a matter ${F}^{(MA)}_E$
and electromagnetic ${F}^{(EM)}_E$
part, although in the
force-free limit the matter part vanishes \cite{McKinney04}.
This power scales with the black hole spin parameter, and in our simulattions,
for $a=0.98$, we obtained the luminosity of the Blandford-Znajek process
on the order of $10^{53}$ erg/s.

A similar order of magnitude is obtained via integration of the neutrino emissivities
over the volume in our simulation.
Also in this case, the results scale with the black hole spin, as the 
conditions in the plasma surrounding the fastly rotating black hole produce mode dense
and hotter disks and hence are brighter in neutrinos.
The emissivities are computed from the balance of nuclear reactions,
which act on partially degenerate and relativistic species: free protons, neutrons,
electrons and positrons. The contributions to the pressure and internal energy
from all these species, which constitute the Fermi gas, are calculated numerically and 
updated consistently at every time step during the dynamical simulation 
(see \cite{Janiuk2007} for the EOS details, and \cite{Janiuk2017} for the details of numerical scheme
and the GR MHD scheme implementation).

\section{Synthesis of heavy elements in the engine under statistical equilibrium}

\begin{center}
\begin{figure}
\includegraphics[width=3.5cm]{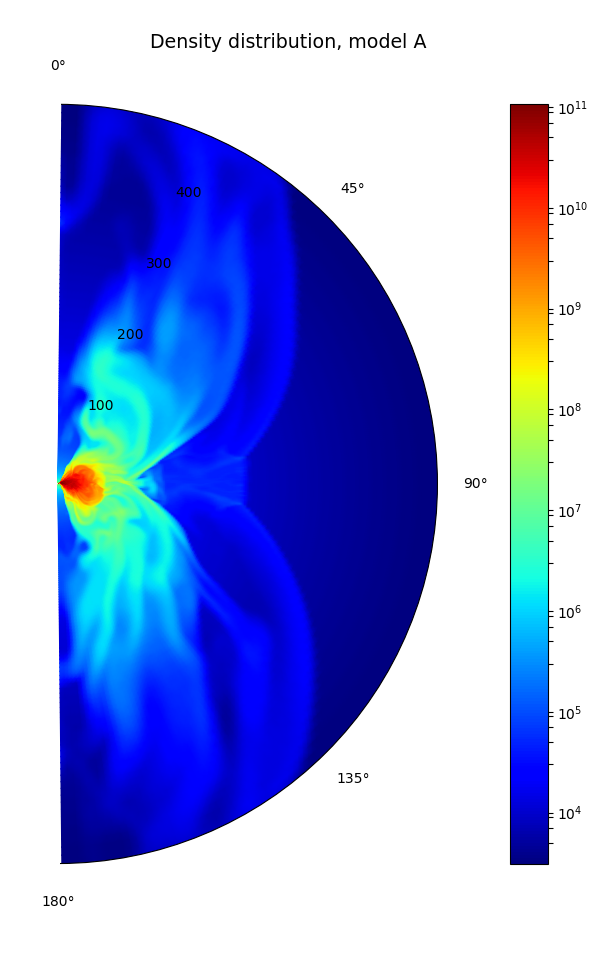}
\includegraphics[width=3.5cm]{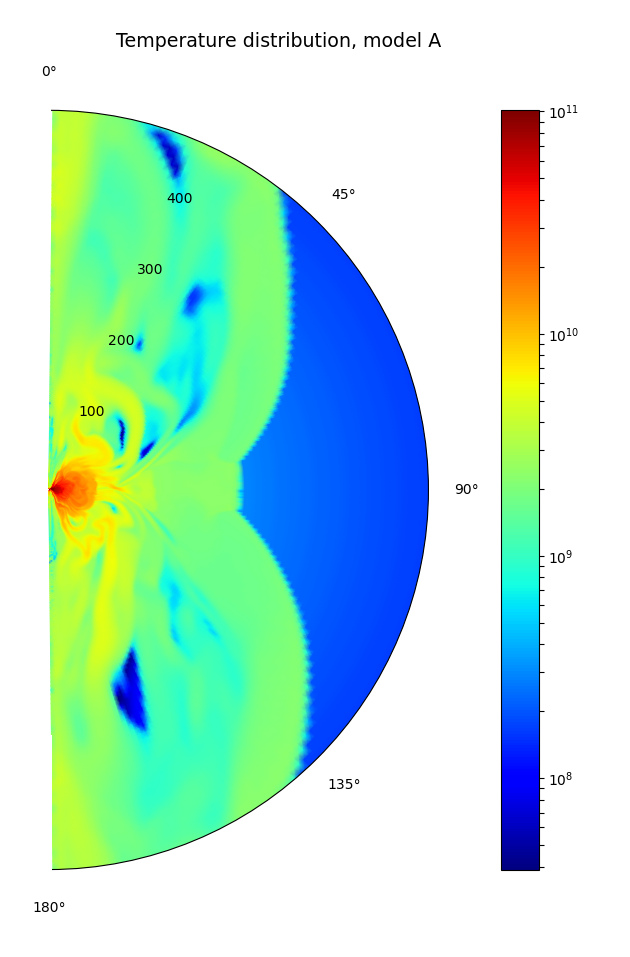}
\caption{Distribution of 
density (left)
and temperature (right)
in the accretion inflow and outflow, up to the distance 
of 500 gravitational radii from the black hole. 
The flow contains a small torus in the GRB engine, and a wind.
The evolved state of the torus is taken at time t=2000 M, after the initial conditions were relaxed. The ouflow material reaches already the outeer boundary, 
but at moderate altitudes.
Global parameters of the model are $M_{\rm BH}=3 M_{\odot}$, $a=0.6$, $M_{\rm disk}=0.1 M_{\odot}$.}
\label{fig:tempA}
\end{figure}
\end{center}

We studied the synthesis of heavy nuclides in the plasma accreting onto
the black hole in the GRB engine, or ejected in the winds from accretion torus.
The physical parameters of the system, such as the black hole mass, 
its spin, and mass of the disk, shape the profiles of temperature and density
in the torus. Exemplary distributions of these 
quentities, within the 500 gravitational radii from the black hole, are shown in Figure \ref{fig:tempA}.
In these calculations, we assume the conditions of nuclear 
statistical equilibrium,
and the number density ratio of protons to neutrons achieves a steady state
value. This so called 'electron fraction'  is then used to establish the
rates of nuclear reactions, which 
together with the tmperature and density profiles on our grid, 
are taken as an iput for the reactions network.
The latter is performed via the postprocessing of the results,
and the currently used computational package is taken from
{\textit http://webnucleo.org}.
The computational 
methods are described in detail in the literature \cite{wallerstein}.

\begin{center}
\begin{figure}
\includegraphics[width=8cm]{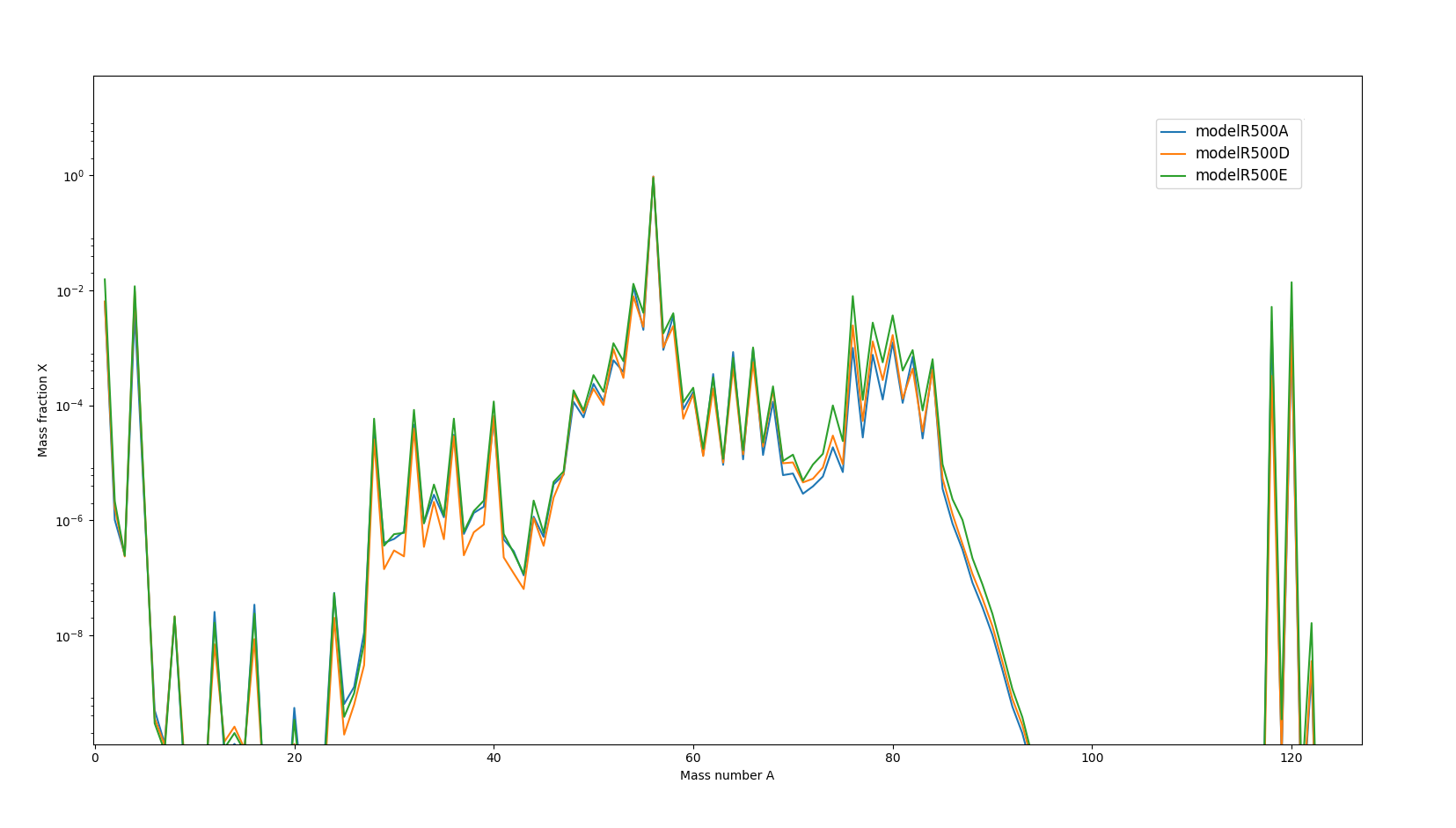}
\caption{Relative mass fraction of nuclei synthesized in the flow
around a black hole in the GRB engine, i.e. torus and its wind.
The parameters of the model are $M_{\rm BH}=3 M_{\odot}$, 
and $M_{\rm disk}=0.1 M_{\odot}$. Three lines mark the results obtained 
for a range of black hole spins,  $a=0.6, 0.8$, and  $a=0.9$.}
\label{fig:nickel}
\end{figure}
\end{center}

The nucleosynthesis results in abundant appearance of lightest isotopes, 
i.e., Helium, Lithium, and Beryllium, and then also the elements
which mass numbers peak around $\sim 80$.
It gives therefore the first peak of nuclides synthesized in the r-process.
The subsequent second and third peaks will be produced 
via neutron capture, in the dynamical ejecta from the disk, at large distances.
In the results presented in Figure \ref{fig:nickel} we observe the
presence of a second peak at $A\sim 120$.

Here, we show the distributions of isotopes that
are produced in the surface layers of accretion disk, and 
within the outflowing gas. The conditions in the nuclear matter
allow for abundant production of nuclei such as 
$^{28}Si$, $^{32}P$, $^{36}Cl$, $^{40}Ca$,
and $^{44}Ti$.
As the Figure show, the amount of $^{56}Ni$, peaks at almost 100\% regardless
of the vallue of the black hole spin.
All the other elements are about 100 times less abundant.
The abundance of elements at around 
$A=80$
($^{76}Cu$, $^{78}Zn$, $^{80}Se$, $^{82}Ge$, $^{84}Ga$),
depends slightly on the value of the black hole spin,
and is the largest for a maximally rotating hole (spin 0.98, model E).
The elements present at around 
$A=120$, 
i.e., $^{118}Kr$, $^{120}Sr$, $^{122}Zr$,
are synthesized also in this model, and their
abundance is not sensitive to the BH spin value.
The model does not predict the third r-process peak at $A\sim200$,
which would require us to go beyond the equilibrium nucleosynthesis and 
follow the dynamical ejecta, with the extrapolation of
the particle trajectories outside the computational domain.
This is postponed to the future work.

\section{Variability of the pulses in short GRB engine due to the magnetic barrier}

In the previous simulations we focused on modeling the structure of torus that
accretes onto the black hole in GRB engine. The detailed microphysics was 
introduced into the dynamical simulations, however these models assumed
 a moderately low magnetic field, normalized with the gas to magnetic pressure
ratio of $\beta=50$, and with an initial poloidal configuration. The magnetic field lines traced the constant density isolines in the torus.
In the last part of our study we present the
study of magnetic field influence on the resulting jet
outflows and their variability. 

Here we adopt a simplified, polytropic equation of state in the gas, 
with the index 
of $4/3$, which is relevant to describe the flow consisting of 
relativistic particles. The degeneracy of species in now neglected.
However, we allow for an extremely high magnetisation 
($\beta=3\times 10^{-3}$). For comparison, we present also the results for
an almost completely thermally dominated flow 
($\beta=3\times 10^{18}$).

\begin{center}
\begin{figure}
\includegraphics[width=5cm]{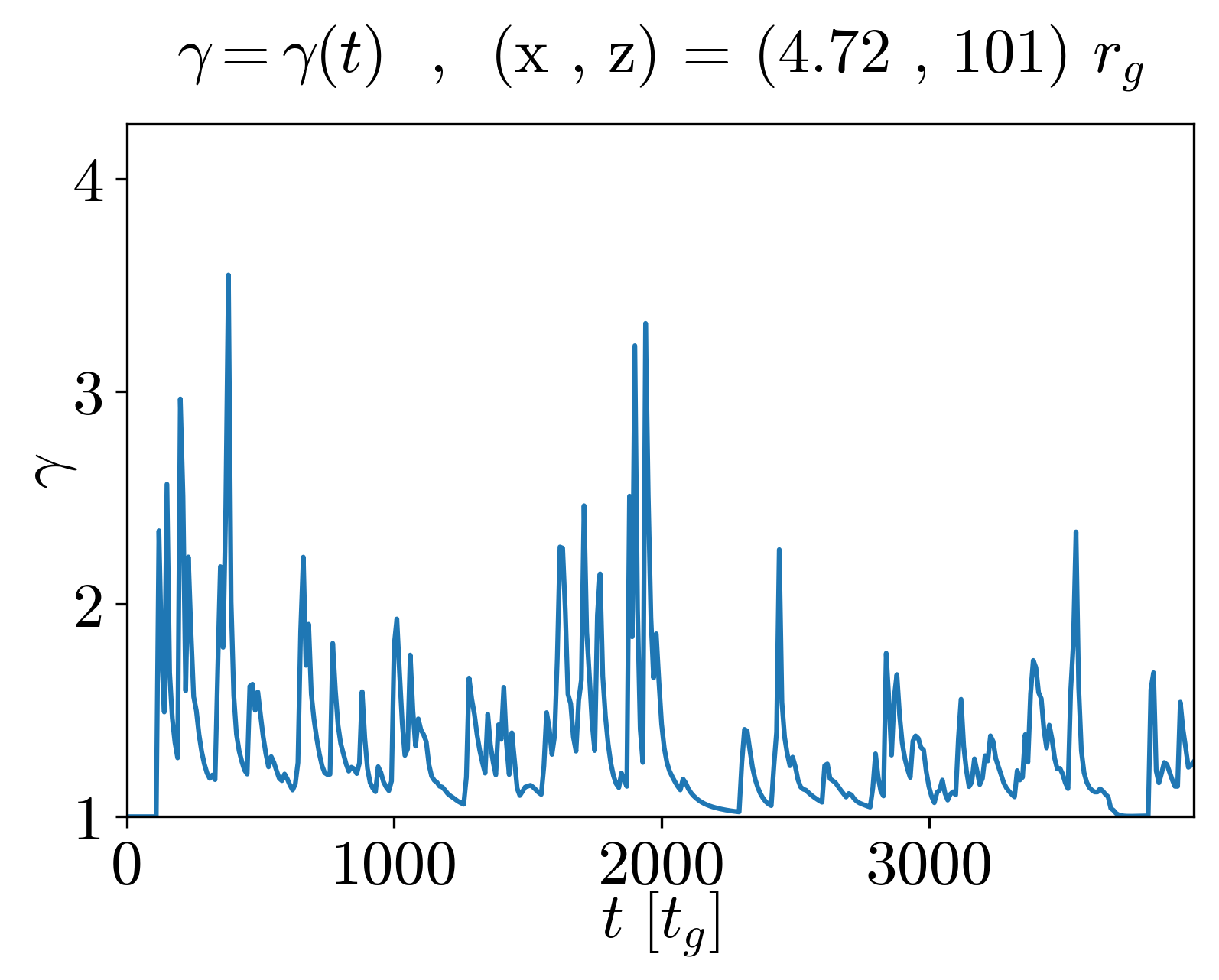}
\includegraphics[width=5cm]{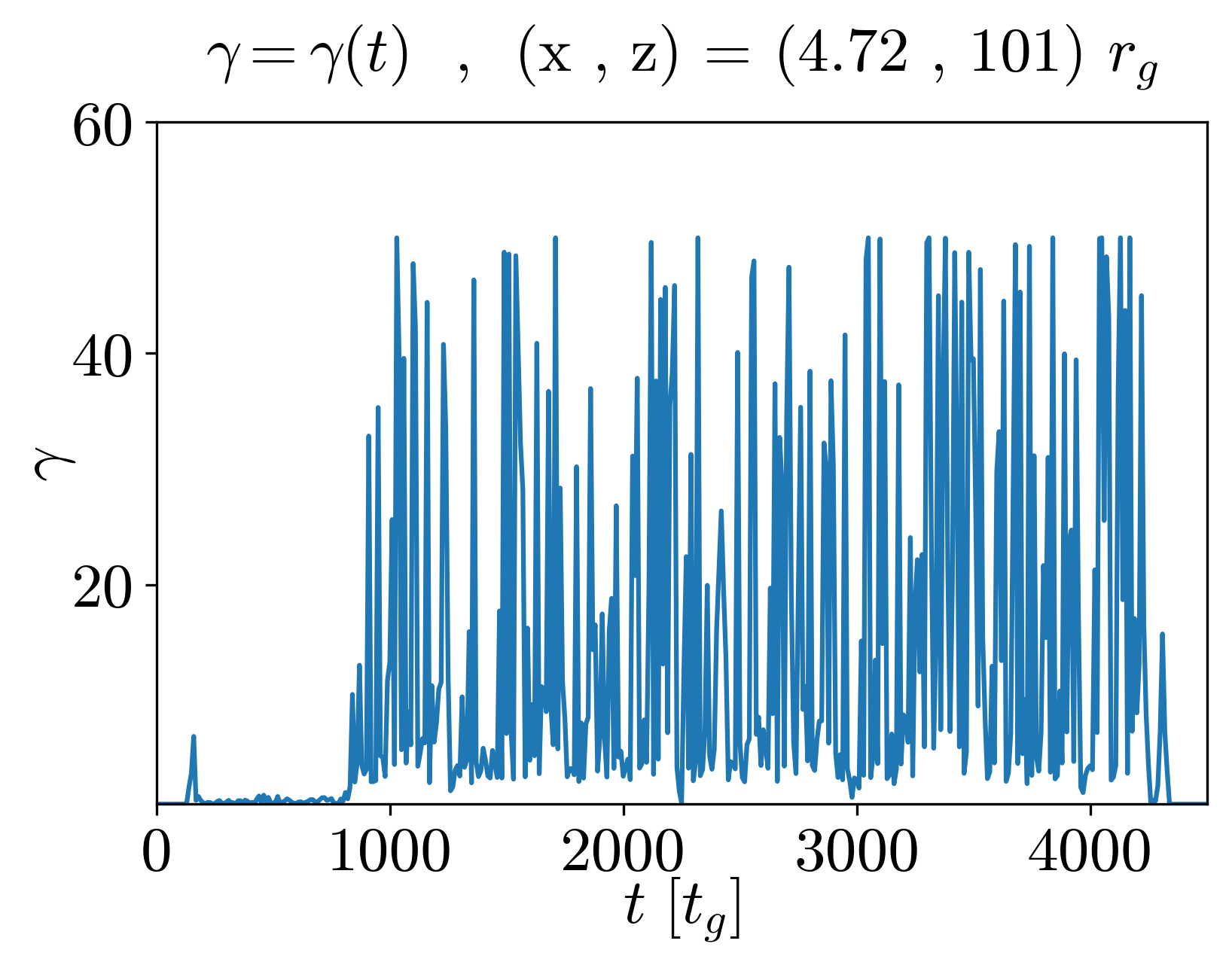}
\caption{
Time profiles of the ejected pulses in the extremely highly magnetized 
(right) and very mildly magnetized (left) central engines. 
The evolved state of the torus is taken at time t=1600 M 
after the initial conditions were relaxed.
The global parameters of the flow are black hole mass 3 $M_{\odot}$ and its 
spin is $a=0.9$.
}
\label{fig:beta}
\end{figure}
\end{center}

The main focus is now given to the time variability of blobs ejected along the
rotation axis of the black hole (a Kerr hole with a spin of $a=0.9$ was 
adopted here), which is quantified by the value of their Lorentz factor.
The thermally dominated and magnetically dominated models presented in
Fig. \ref{fig:beta} yield the $\Delta t_{\rm blobs}$ on the order of
$4.9\times 10^{-3}$ sec, and $5.9\times 10^{-4}$ sec, respectively.
Also, the lower $\beta$ parameter produces higher Lorentz factors of the jets.
This result provides a quantitative way to estimate the magnitude of
magnetic field of the progenitor, basing on the observed profiles
of the GRB lightcurve.
We note here that, due to the significant acceleration our outflow achieves, it was necessary to consider a gamma barrier and test different values to exclude any implication on the launched jet variability. The selected value is $\gamma_{\rm max} = 50$ and is common for all models.
Going beyond this limitation is the subject of our future study.

\section{Conclusions}

A reliable model of the process leading to the high energy emission in the
GRB jets requires detailed modeling of the central engine that is at the base 
of GRB and drives the jets.
This is done in our simulations by self-consistenly solving the
equations of General Relativistic magnetohydrodynamics.
In this way we are able to quantify the conditions sufficient for 
the appearance of magnetically driven jets, and also for the 
uncollimated winds (outflows) from the engine.
We also argue for the neutrino-antineutrino annihillation process being comparable in its efficiency of driving the jets to the well-established 
Blandford-Znajek mechanism.

By postprocessing of our results, we are aiming to quantify the amouns of heavy nuclei produced in te GRB engine and its outflows ejecta.
In addition to the a possible detection of heavy nuclei and their radioactive 
decay signal in the afterglows, the variability of prompt GRB 
emission is the main and direct observable.
Both these quantities link our modeling to the
observable GRB sources.

{\bf Acknowledgements}
We acknowledge partial support from grant DEC-2012/05/E/ST9/03914 and 
DEC-2016/23/B/ST9/03114
awarded by the Polish National Science Center.

\end{document}